\documentclass[aps,10pt]{revtex4}
\def\theequation{\arabic{section}.\arabic{equation}}
\usepackage{subfigure}
\usepackage{xcolor}
\usepackage{mathrsfs}
\usepackage{graphicx}
\usepackage[colorlinks=true,linkcolor=blue,citecolor=magenta,urlcolor=blue]{hyperref}
\usepackage{amssymb, bm}
\usepackage{amsmath, amsthm}
\usepackage{epstopdf}
\usepackage{hyperref}
\usepackage{enumerate}
\usepackage{longtable}
\usepackage{amsmath}

\usepackage{amsmath}  
\usepackage{amsfonts} 
\usepackage{graphicx}

\newcommand{\be}{\begin{equation}}
\newcommand{\ee}{\end{equation}}

\newcommand{\ms}{\medskip}

\begin{document}
\def\theequation{\arabic{section}.\arabic{equation}}

\title{
Clothoid helices obtained via the Lie-Darboux method}  

\author{H.C. Rosu}
\email{hcr@ipicyt.edu.mx; ORCID: 0000-0001-5909-1945} 
\affiliation{Instituto Potosino de Investigaci\'on Cient\'{\i}fica y Tecnol\'ogica, Camino a la Presa San Jos\'e 2055,
Col. Lomas 4a Secci\'on, San Luis Potos\'{\i}, 78216 S.L.P., Mexico}

\author{J. de la Cruz}
\email{josue.delacruz@ipicyt.edu.mx; ORCID: 0000-0001-5943-5752}
\affiliation{Instituto Potosino de Investigaci\'on Cient\'{\i}fica y Tecnol\'ogica, Camino a la Presa San Jos\'e 2055,
Col. Lomas 4a Secci\'on, San Luis Potos\'{\i}, 78216 S.L.P., Mexico}

\author{P. Lemus-Basilio}
\email{paola.lemus@ipicyt.edu.mx; ORCID: 0000-0002-7364-3643}
\affiliation{Instituto Potosino de Investigaci\'on Cient\'{\i}fica y Tecnol\'ogica, Camino a la Presa San Jos\'e 2055,
Col. Lomas 4a Secci\'on, San Luis Potos\'{\i}, 78216 S.L.P., Mexico}


\bigskip
\bigskip
\begin{abstract}
The clothoid helices that have both curvature and torsion directly proportional to the arclength 
are obtained via the Lie-Darboux method and analyzed in some detail. Shifted counterparts are also introduced and studied within the same framework.

\medskip


\end{abstract}


\maketitle

\section{Introduction}\label{sec:1}
\setcounter{equation}{0}

In the late 19th century, Lie and Darboux obtained one remarkable result of differential geometry of regular curves in three-dimensional Euclidean space, namely that they are characterized by a particular case of the Riccati equation with the curvature $\kappa$ and torsion $\tau$ as coefficients of the equation as follows 
\begin{equation}\label{jph1}
	\frac{dw}{ds}=-i\kappa(s) w+i\frac{\tau(s)}{2}w^2-i\frac{\tau(s)}{2}~,
\end{equation}
where the independent variable $s$ is the arc length of the curve. 

\ms

The helices are an important case of space curves that are characterized by $\kappa=k\,\tau$, where $k$ is an arbitrary real constant.
This property allows separation of variables and partial fractions decomposition, leading to rational solutions of (\ref{jph1}).
However, the Lie-Darboux (LD) result, although mentioned in all classical textbooks in
differential geometry \cite{bEisen1909,bScheff1923,bStruik1961}, has not been used in the literature since many decades, perhaps because many Riccati solutions can be obtained only numerically and also it is not so straightforward to go from the Riccati solution $w$ to the intrinsic parametric equations of the curve.
Recently, the first and last authors revisited the LD method, finding that it yields not only the Riccati equation (\ref{jph1}), 
but also a counterpart  
with opposite sign of torsion \cite{ph2023}.

\ms

Here we study the clothoid helices that have $\kappa(s)$ and $\tau(s)$ directly proportional to the arclength using the Lie-Darboux method. 
They  
represent a three-dimensional generalization of the clothoid spirals, a name proposed by Ces\`aro around 1890 for what are also known as the Cornu spirals, which are 
the two-dimensional (zero torsion) curves  having $\kappa(s)$ directly proportional to the arclength.
The literature on clothoid helices covers only a few papers. Frego \cite{Frego2022} presented an important contribution in which a Lie-group approach of the Frenet-Serret system on such curves is  
undertaken and various geometric integrators are applied. A few additional papers \cite{Li2001,Harary,Casati} are also briefly reviewed therein. We also note that L\'opez and Weber \cite{LW} have constructed minimal surfaces of Bj\" orling type based on clothoid helices.  

\ms

The organization of the paper is the following. In Section II, we present the basics of the LD method. In Section III, two types of clothoid helices
are obtained by employing the LD method. In Section IV, we deal with a shifted-arclength case, which is a slightly more general case than the previous ones.
Finally, Section V contains a few conclusions.

\section{The LD method}\label{sec:1}

The LD method consists of three steps:

\ms

(i). Firstly, one requires a rational solution of (\ref{jph1}).

In the case of the clothoid helices, the Riccati solution has the form
\begin{equation}\label{jph2}
	w(s)=\frac{w_1\exp\left(i\frac{\sqrt{k^2+1}}{2}\frac{ s^2}{c^2}\right)+w_2}{\exp\left(i\frac{\sqrt{k^2+1}}{2}\frac{ s^2}{c^2}\right)+1} ~,
\end{equation}
where $w_1=k+\sqrt{k^2+1}$ and $ w_2=k-\sqrt{k^2+1}$.

%
%
%


\ms

\noindent (ii). In the second step, one uses the following formulas for the components $\alpha_i$ of the unit tangent vector of a spatial curve
	\begin{align}
		\alpha_1&=\frac{f_1^2-f_2^2-f_3^2+f_4^2}{2 (f_1 f_4-f_2 f_3)}~,\nonumber\\
		\alpha _2&=i\frac{ \left(f_1^2+f_2^2-f_3^2-f_4^2\right)}{2 (f_1 f_4-f_2 f_3)}~, \label{Schef}\\
		\alpha_3&=\frac{ \left(f_3f_4-f_1f_2\right)}{ (f_1f_4-f_2 f_3)}~,\nonumber
	\end{align}
	where the pairs of functions $(f_1,f_2)$ and $(f_3,f_4)$ are those in the numerator and denominator, respectively, of the rational Riccati solution (\ref{jph2}) written in the general form
	\begin{equation}\label{jph5}
	w(s)=\frac{{\rm C}f_1+f_2}{{\rm C}f_3+f_4} ~,	
	\end{equation}
		where ${\rm C}$ is an arbitrary constant.

\ms

\noindent	(iii). The final step consists of computing and plotting the Cartesian coordinates of the helix according to the integrals
	\begin{align}
		x(s)=&\int^s\alpha_1(\sigma)d\sigma~, \nonumber\\
		y(s)=&\int^s\alpha_2(\sigma)d\sigma~,\\
		z(s)=&\int^s\alpha_3(\sigma)d\sigma~.\nonumber
	\end{align}

Note that the clothoid Riccati solution (\ref{jph2}) takes the form (\ref{jph5}) for ${\rm C}=1$. This allows one to
identify the functions $f_i$ in all possible permutation ways from (\ref{jph2}).

\section{The clothoid helices}
As already noted above, we will consider $\kappa(s)= ks/c^2$ and $\tau(s)= s/c^2$, i.e., the case of clothoid helices with the quotient $\kappa/\tau=k$.
The parameter $c$  serves as a supplementary dilation parameter.

\ms

There are four possible ways to choose the set $\{f_1,f_2,f_3,f_4\}$:
\begin{equation}	\label{jph6-1}
	{\bf 1.} \quad f_1=w_1 e^{i\frac{\sqrt{k^2+1}}{2}\frac{ s^2}{c^2}}~,~~f_2=w_2~,~~f_3=e^{i\frac{\sqrt{k^2+1}}{2}\frac{ s^2}{c^2}}~,~~f_4=1 ~,
\end{equation}
\begin{equation}\label{jph6-2}
	{\bf 2.} \quad f_1=w_2~,~~f_2=w_1e^{i\frac{\sqrt{k^2+1}}{2}\frac{ s^2}{c^2}}~,~~f_3=1~,~~f_4=e^{i\frac{\sqrt{k^2+1}}{2}\frac{ s^2}{c^2}}~, 
	\end{equation}
 \begin{equation}\label{jph6-3}
	{\bf 3.} \quad f_1=w_1e^{i\frac{\sqrt{k^2+1}}{2}\frac{ s^2}{c^2}}~,~~f_2=w_2~,~~
	f_3=1~,~~f_4=e^{i\frac{\sqrt{k^2+1}}{2}\frac{ s^2}{c^2}}~,
\end{equation}
 \begin{equation}\label{jph6-4}
		{\bf 4.} \quad f_1=w_2~,~~f_2=w_1e^{i\frac{\sqrt{k^2+1}}{2}\frac{ s^2}{c^2}}~,~~
		f_3=e^{i\frac{\sqrt{k^2+1}}{2}\frac{ s^2}{c^2}}~,~~f_4=1~.
	\end{equation} 

The last two sets do not yield closed-form analytical results and will not be considered further here.

\ms

{\bf Case 1}

\ms

In the first case, the $\alpha_i$ components are obtained in the form:
\begin{align}
	\alpha_1(s)&=k \Bigg[\cos \left(\frac{\sqrt{k^2+1}}{2}\frac{ s^2}{c^2}\right)+i\frac{k}{\sqrt{k^2+1}} \sin \left(\frac{\sqrt{k^2+1}}{2}\frac{ s^2}{c^2}\right)\Bigg]~,\nonumber\\
	\alpha _2(s)&=k \Bigg[-\sin \left(\frac{\sqrt{k^2+1}}{2}\frac{ s^2}{c^2}\right)+i\frac{k}{\sqrt{k^2+1}} \cos \left(\frac{\sqrt{k^2+1}}{2}\frac{ s^2}{c^2}\right)\Bigg]~,\\
	\alpha_3(s)&=\frac{1}{\sqrt{k^2+1}}\nonumber~,
\end{align}
which satisfy $\alpha_1^2+\alpha_2^2+\alpha_3^2=1$. The coordinates on the helix are given by
	\begin{align}
	x_1(s)=&\int^s\alpha_1(\sigma)d\sigma=\frac{\sqrt{\pi}ck}{(k^2+1)^{\frac{1}{4}}}\Bigg[C\left(\frac{(k^2+1)^{\frac{1}{4}}}{\sqrt{\pi}c}s\right)+
           i\frac{k}{(k^2+1)^{\frac{1}{2}}}S\left(\frac{(k^2+1)^{\frac{1}{4}}}{\sqrt{\pi}c}s\right)\Bigg]~,	\nonumber\\
	y_1(s)=&\int^s\alpha_2(\sigma)d\sigma=\frac{\sqrt{\pi}ck}{(k^2+1)^{\frac{1}{4}}}\Bigg[-S\left(\frac{(k^2+1)^{\frac{1}{4}}}{\sqrt{\pi}c}s\right)+
           i\frac{k}{(k^2+1)^{\frac{1}{2}}}C\left(\frac{(k^2+1)^{\frac{1}{4}}}{\sqrt{\pi}c}s\right)\Bigg]~,	 	\label{c-2-y1}\\
	z_1(s)=&\int^s\alpha_3(\sigma)d\sigma=\frac{s}{\sqrt{k^2+1}}~,\nonumber		
	\end{align}
or, in a column vector notation,
\begin{equation}\label{3-9}
\vec{{\cal C}}_1(s)=\left(
\begin{array}{c}
	x_1\\
	y_1\\
	z_1
\end{array}
\right)
=
\left(
\begin{array}{c}
	\int^s\alpha_1(\sigma)d\sigma\\ 
	\int^s\alpha_2(\sigma)d\sigma\\ 
	\int^s\alpha_3(\sigma)d\sigma\\ 
\end{array}
\right)
=
\left(
\begin{array}{c}
	\frac{\sqrt{\pi}ck}{(k^2+1)^{\frac{1}{4}}}C\left(\frac{(k^2+1)^{\frac{1}{4}}}{\sqrt{\pi}c}s\right)\\
	-\frac{\sqrt{\pi}ck}{(k^2+1)^{\frac{1}{4}}}S\left(\frac{(k^2+1)^{\frac{1}{4}}}{\sqrt{\pi}c}s\right)\\
	\frac{s}{\sqrt{k^2+1}}
\end{array}
\right)
+i
\left(
\begin{array}{c}
	\frac{\sqrt{\pi}ck^2}{(k^2+1)^{\frac{3}{4}}}S\left(\frac{(k^2+1)^{\frac{1}{4}}}{\sqrt{\pi}c}s\right)\\
	\frac{\sqrt{\pi}ck^2}{(k^2+1)^{\frac{3}{4}}}C\left(\frac{(k^2+1)^{\frac{1}{4}}}{\sqrt{\pi}c}s\right)\\
	0
\end{array}
\right)~.
\end{equation}

\begin{figure}[htb]\centering
		\subfigure[~ $Re(x_1,y_1,z_1); s\in (-\sqrt{50},+\sqrt{50})$]{\includegraphics[width=0.45\linewidth] {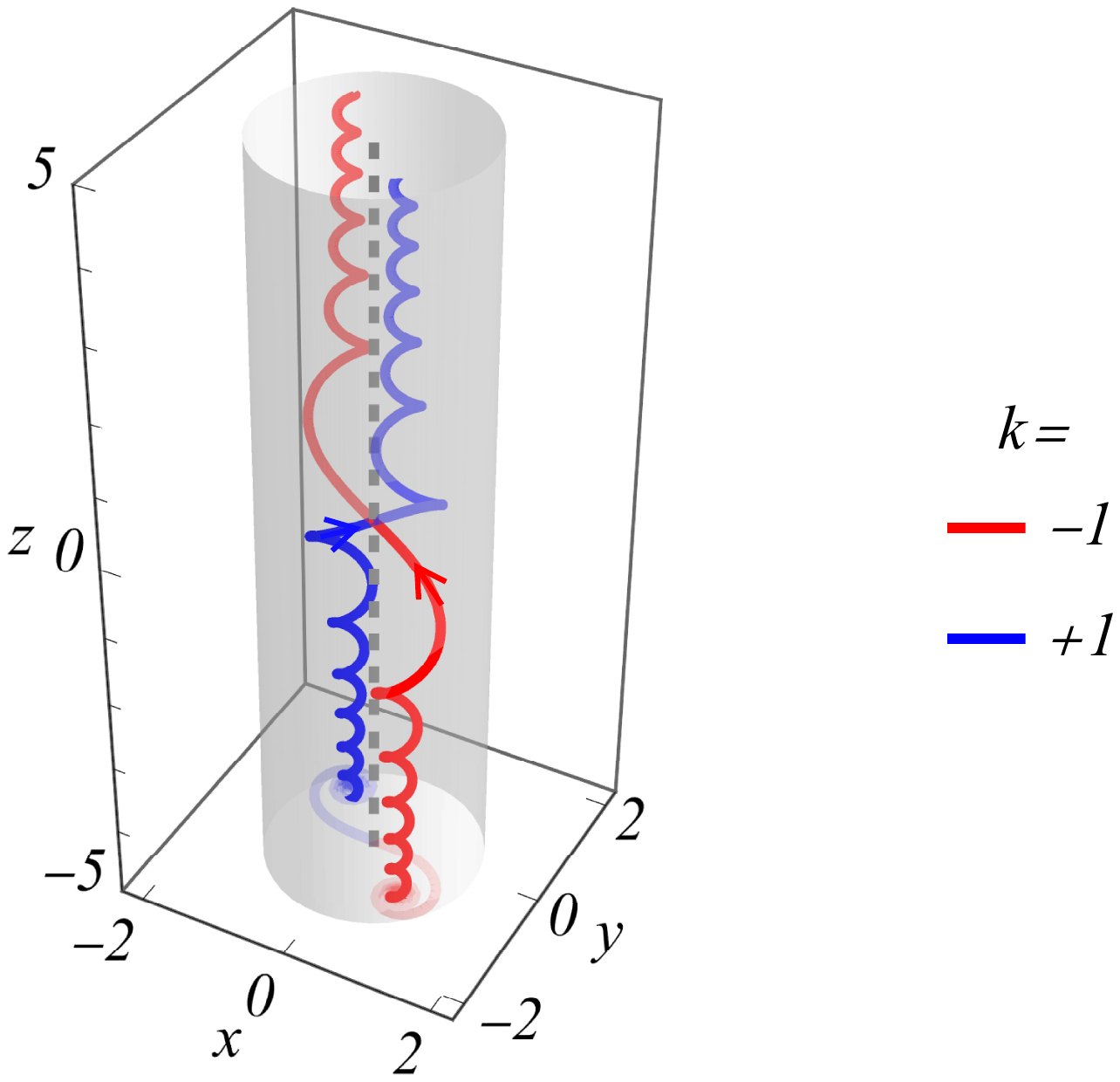}} 
		\subfigure[~ $Re(x_1,y_1,z_1); s\in (-\sqrt{125},+\sqrt{125})$]{\includegraphics[width=0.45\linewidth]   {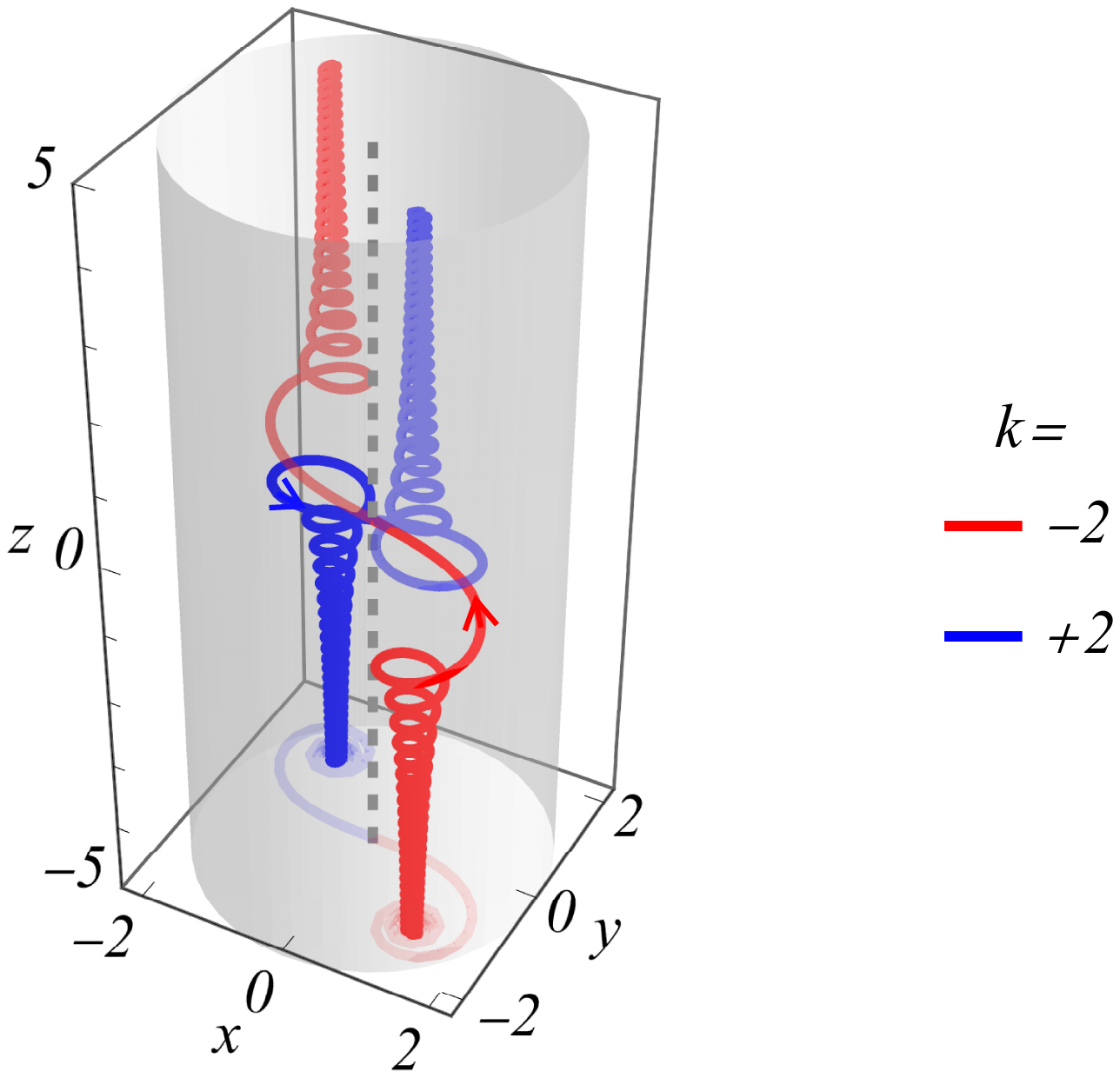}}    
		\caption{The clothoid helix $\vec{{\cal C}}_1$ from (\ref{3-9}) for $c=1$ and $k=\pm 1$ (a) and $k=\pm 2$ (b).}\label{fc1}
	\end{figure}

We observe that the coordinates $x_1$ and $y_1$ are complex-valued quantities while the $z_1$ coordinate is real. Thus, only the real part describes a clothoid helix, whereas the imaginary part describes a clothoid spiral. 
Plots of the helix part in (\ref{3-9}) are given in Fig.~\ref{fc1} for $c=1$ and $k=\pm 1$ and $k=\pm 2$. 

\ms

The coordinates of the foci of this clothoid helix can be obtained using the asymptotic property of the Fresnel integrals, $C(\pm\infty)=S(\pm\infty)=\pm\frac{1}{2}$, 
	\begin{equation}\label{foci-1}	
       x_{1,F}(s\to\pm\infty)=\pm\frac{ck\sqrt{\pi}}{2(k^2+1)^{\frac{1}{4}}}~,~\quad y_{1,F}(s\to\pm\infty)=
       \mp\frac{ck\sqrt{\pi}}{2(k^2+1)^{\frac{1}{4}}}~. 
	\end{equation}
From (\ref{foci-1}), one can readily deduce that the two foci lie on the second bisector.

\ms

{\bf Case 2}

\ms

	For the second case, denoting by $\tilde{\alpha}_i$ the tangential components, it follows directly from (\ref{Schef}) that 
\begin{equation}\label{talpha2}
\tilde{\alpha}_1=\alpha_1(s)~, \quad \tilde{\alpha}_2=-\alpha_2(s)~, \quad \tilde{\alpha}_3=-\alpha_3(s)~,
\end{equation}
\begin{equation}\label{3-9b}
\vec{{\cal C}}_2(s)=\left(
\begin{array}{c}
	x_1\\
	-y_1\\
	-z_1
\end{array}
\right)
=
\left(
\begin{array}{c}
	\frac{\sqrt{\pi}ck}{(k^2+1)^{\frac{1}{4}}}C\left(\frac{(k^2+1)^{\frac{1}{4}}}{\sqrt{\pi}c}s\right)\\
	\frac{\sqrt{\pi}ck}{(k^2+1)^{\frac{1}{4}}}S\left(\frac{(k^2+1)^{\frac{1}{4}}}{\sqrt{\pi}c}s\right)\\
	-\frac{s}{\sqrt{k^2+1}}
\end{array}
\right)
+i
\left(
\begin{array}{c}
	\frac{\sqrt{\pi}ck^2}{(k^2+1)^{\frac{3}{4}}}S\left(\frac{(k^2+1)^{\frac{1}{4}}}{\sqrt{\pi}c}s\right)\\
	-\frac{\sqrt{\pi}ck^2}{(k^2+1)^{\frac{3}{4}}}C\left(\frac{(k^2+1)^{\frac{1}{4}}}{\sqrt{\pi}c}s\right)\\
	0
\end{array}
\right)~.
\end{equation}

\begin{figure}[htb]\centering
        \subfigure[~ $Re(x_2,y_2,z_2); s\in (-\sqrt{50},+\sqrt{50})$]{\includegraphics[width=0.45\linewidth]  {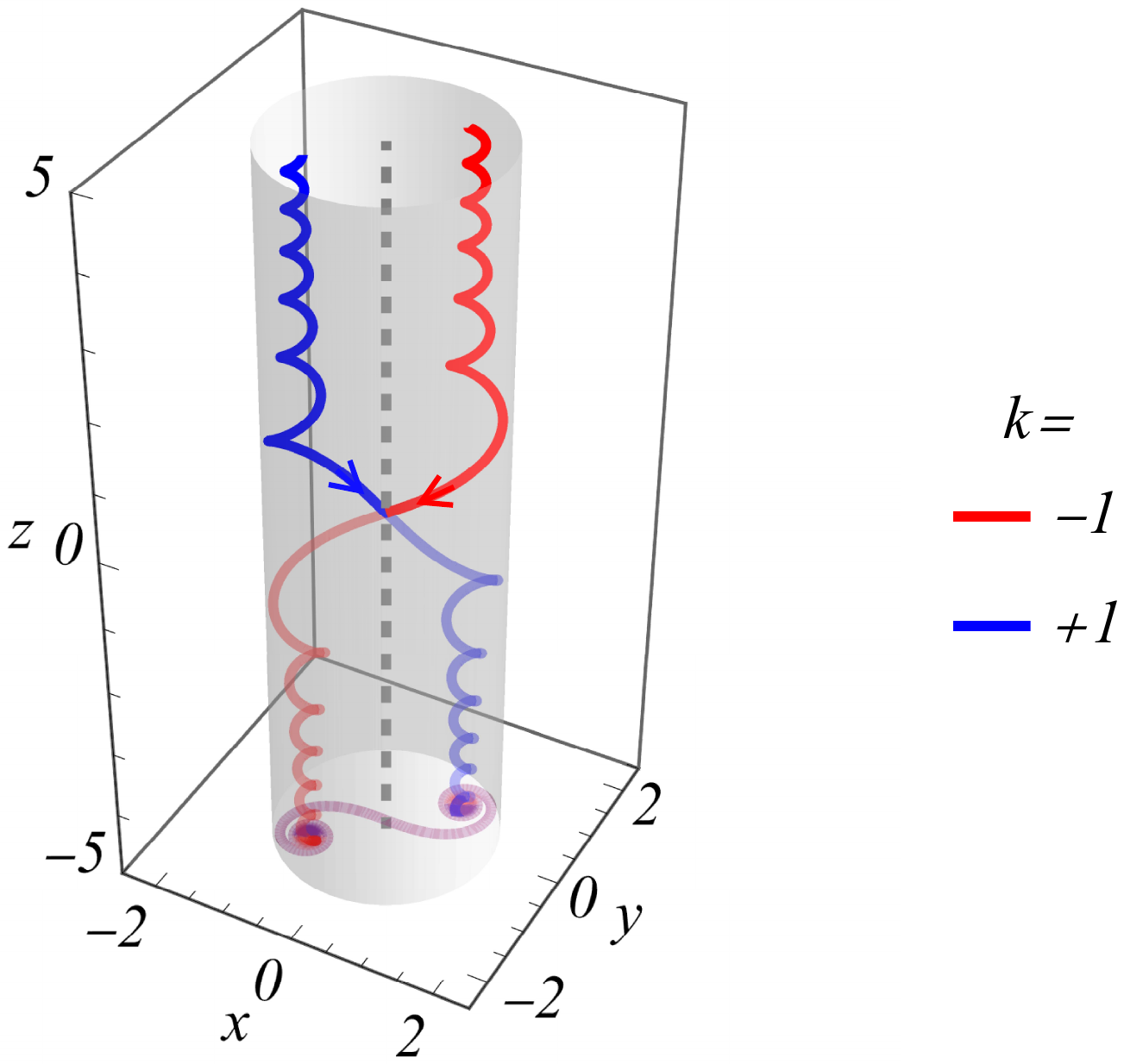}}  
        \subfigure[~ $Re(x_2,y_2,z_2); s\in (-\sqrt{125},+\sqrt{125})$]{\includegraphics[width=0.45\linewidth]  {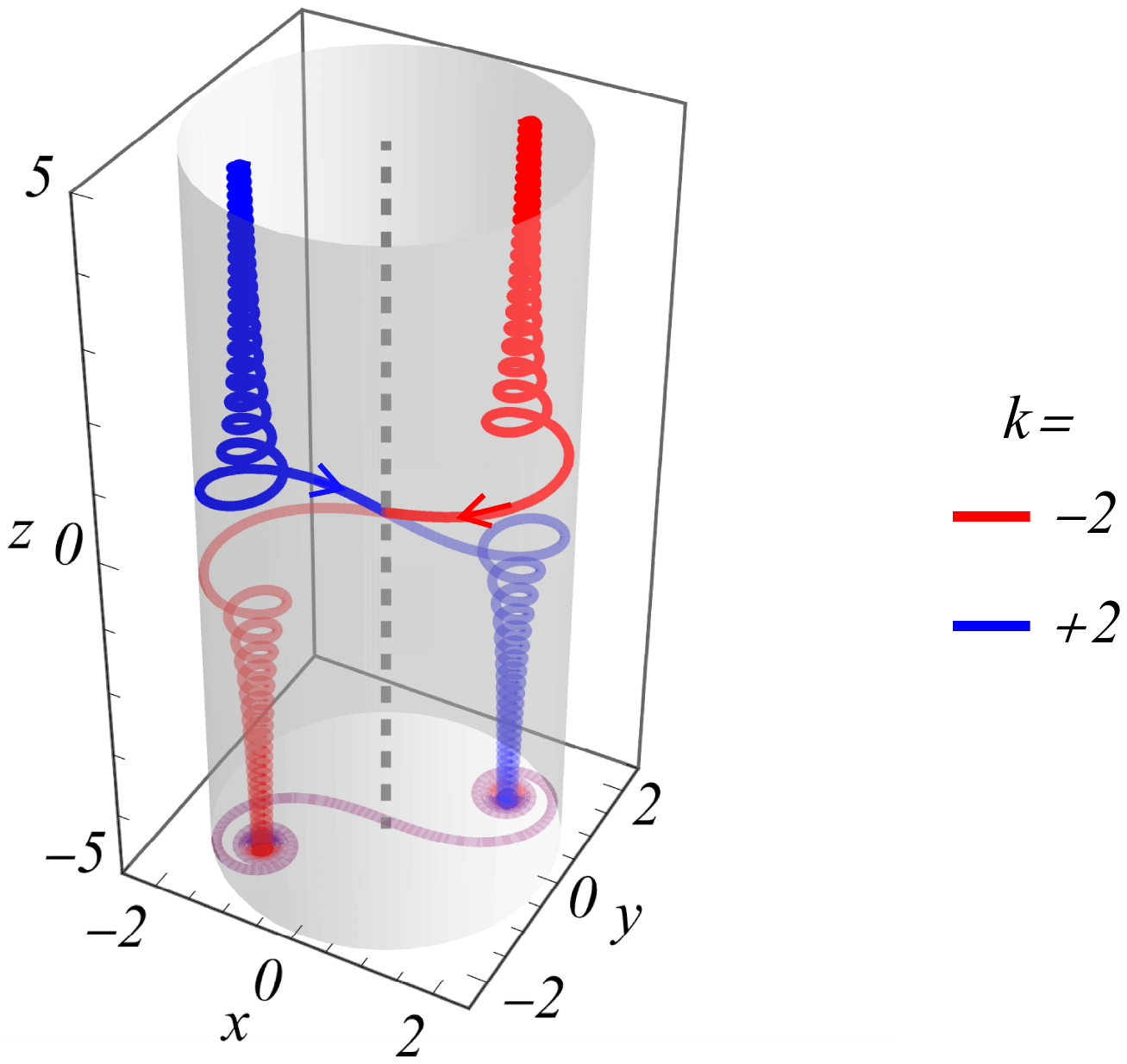}} 
		\caption{Same as in Fig. 1 for the clothoid helix $\vec{{\cal C}}_2$ from (\ref{3-9b}).}\label{fc2}
	\end{figure}

Plots analogous to those in Fig.~\ref{fc1} are provided in Fig.~\ref{fc2} for $\vec{{\cal C}}_2$. 
This clothoid helix has its foci on the first bisector as can be seen from the coordinates given by 

	\begin{equation}\label{foci2}
x_{2,F}(s\to\pm\infty)=y_{2,F}(s\to\pm\infty)=\pm\frac{ck\sqrt{\pi}}{2(k^2+1)^{\frac{1}{4}}}~. 
	\end{equation}

\section{The $\delta$-shifted clothoid helices}

A more general case than those considered in the previous section is obtained by taking $\kappa(s)=\tau(s)=\frac{s}{c^2}+\delta$ 
where $\delta$ is a constant shift parameter. For simplicity, we restrict to the case $k=1$, for which the Riccati solution takes the rational form
\begin{equation}\label{delta1}
	w(s;\delta,c)=\frac{(1+\sqrt{2}) e^{\frac{i s^2}{\sqrt{2} c^2}+\frac{i \sqrt{2} \delta  s}{c^2}}
+1-\sqrt{2}}{e^{\frac{i s^2}{\sqrt{2} c^2}+\frac{i \sqrt{2} \delta  s}{c^2}}+1}
\end{equation}
and the set $\{f_i\}$ corresponding to the case {\bf 1} reads

\begin{equation}\label{delta2}
	f_1=w_1e^{\frac{i s^2}{\sqrt{2} c^2}+\frac{i \sqrt{2} \delta  s}{c^2}},~~f_2=w_2,~~
	f_3=e^{\frac{i s^2}{\sqrt{2} c^2}+\frac{i \sqrt{2} \delta  s}{c^2}},~~f_4=1~.
\end{equation}

The $\alpha_i$ components of the unit tangent vector are 
	\begin{align}
		\alpha_{1\delta}&=\cos \left(\frac{s (2 \delta +s)}{\sqrt{2} c^2}\right)+i\frac{1}{\sqrt{2}}\sin \left(\frac{s (2 \delta +s)}{\sqrt{2} c^2}\right)\nonumber\\
		\alpha _{2\delta}&=-\sin \left(\frac{s (2 \delta +s)}{\sqrt{2} c^2}\right)+i\frac{1}{\sqrt{2}} \cos \left(\frac{s (2 \delta +s)}{\sqrt{2} c^2}\right)\\
		\alpha_{3\delta}&=\frac{1}{\sqrt{2}}~.\nonumber
	\end{align}
With these $\alpha$'s, the $\delta$-chlotoid helices for this case are obtained in the form  
\begin{equation}\label{3-9c}
\vec{{\cal C}}_{1,\delta}(\tilde{s})=\left(
\begin{array}{c}
	x_{1\delta}\\
	y_{1\delta}\\
	z_{1\delta}
\end{array}
\right)
=
\left(
\begin{array}{c}
	\frac{\sqrt{\pi} c }{2^{1/4}}{\cal F}_1(\tilde{s})\\
	-\frac{\sqrt{\pi} c }{2^{1/4}}{\cal F}_2(\tilde{s})\\
	\frac{\tilde{s}-\delta}{\sqrt{2}}
\end{array}
\right)
+i
\left(
\begin{array}{c}
	\frac{\sqrt{\pi}c}{2^{1/4}}{\cal F}_2(\tilde{s})\\
	\frac{\sqrt{\pi}c}{2^{1/4}}{\cal F}_1(\tilde{s})\\
	0
\end{array}
\right)~,
\end{equation}
where 
\begin{equation}\label{F1}
{\cal F}_1(\tilde{s})=\cos\left(\frac{\delta ^2}{\sqrt{2} c^2}\right)C\left(\frac{\sqrt[4]{2}\tilde{s}}{\sqrt{\pi }c}\right) 
+\sin\left(\frac{\delta ^2}{\sqrt{2} c^2}\right)S\left(\frac{\sqrt[4]{2}\tilde{s}}{\sqrt{\pi}c}\right) 
\end{equation}
\begin{equation}\label{F2}
{\cal F}_2(\tilde{s})= -\sin\left(\frac{\delta ^2}{\sqrt{2} c^2}\right) C\left(\frac{\sqrt[4]{2}\tilde{s}}{\sqrt{\pi }c}\right) +                                                                                      
\cos\left(\frac{\delta ^2}{\sqrt{2} c^2}\right)S\left(\frac{\sqrt[4]{2}\tilde{s}}{\sqrt{\pi}c}\right)  
\end{equation}
and $\tilde{s}=s+\delta$.

Using again $C(\pm\infty)=S(\pm\infty)=\pm\frac{1}{2}$, the locations of the $\delta$-foci 
are the following
	\begin{equation}\label{poles_1}
		x_{1\delta,F}(\tilde{s}\to\pm\infty)=\pm\frac{c\sqrt{\pi}}{2^{5/4}}\left[\cos\left(\frac{\delta^2}{\sqrt{2}c^2}\right)+\sin\left(\frac{\delta^2}
{\sqrt{2}c^2}\right)\right]~,\,\,
		y_{1\delta,F}(\tilde{s}\to\pm\infty)=\pm\frac{c\sqrt{\pi}}{2^{5/4}}\left[\sin\left(\frac{\delta^2}{\sqrt{2}c^2}\right)-\cos\left(\frac{\delta^2}
{\sqrt{2}c^2}\right)\right]~.
	\end{equation}

\begin{figure}[htb]\centering
{\includegraphics[width=0.362\linewidth]  {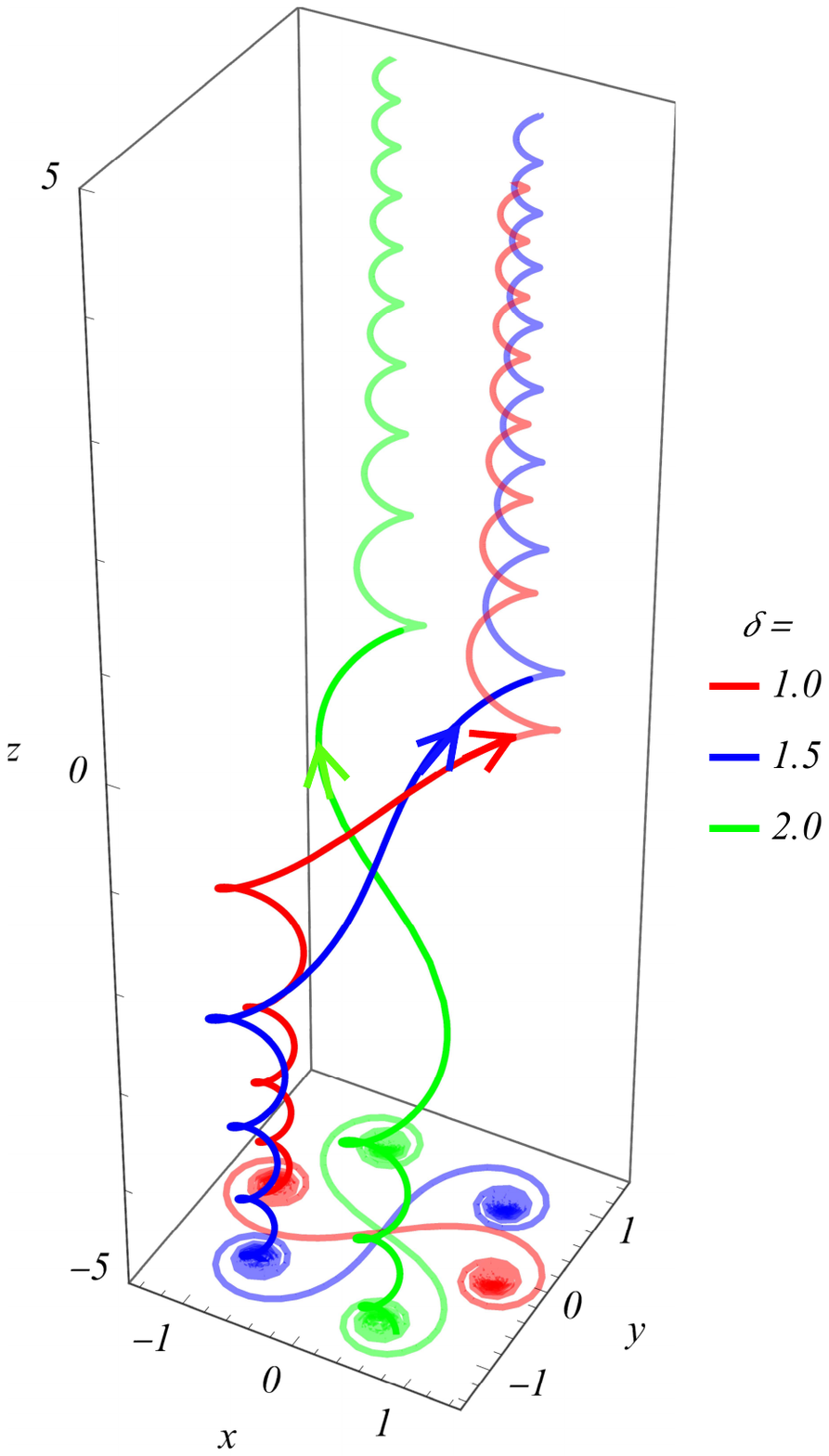}} 
		\caption{$\delta$-shifted clothoid helices $\vec{{\cal C}}_{1,\delta}(\tilde{s})$ for three different values of $\delta$ and $c=1$.}\label{fdeltac1}
	\end{figure}

Plots of $\vec{{\cal C}}_{1,\delta}(\tilde{s})$ for a few values of the $\delta$ shift are displayed in Fig.~\ref{fdeltac1}. 

\ms

For the case {\bf 2}, one obtains:
\begin{equation}\label{talpha2d}
\tilde{\alpha}_{1\delta}=\alpha_{1\delta}(s)~, \quad \tilde{\alpha}_{2\delta}=-\alpha_{2\delta}(s)~, \quad \tilde{\alpha}_{3\delta}=-\alpha_{1\delta}(s)~,
\end{equation}
which yields to the $\delta$-clothoid helix
\begin{equation}\label{3-9c}
\vec{{\cal C}}_{2,\delta}(\tilde{s})=\left(
\begin{array}{c}
	x_{1\delta}\\
	-y_{1\delta}\\
	-z_{1\delta}
\end{array}
\right)
=
\left(
\begin{array}{c}
	\frac{\sqrt{\pi} c }{2^{1/4}}{\cal F}_1(\tilde{s})\\
	\frac{\sqrt{\pi} c }{2^{1/4}}{\cal F}_2(\tilde{s})\\
	-\frac{\tilde{s}-\delta }{\sqrt{2}}
\end{array}
\right)
+i
\left(
\begin{array}{c}
	\frac{\sqrt{\pi}c}{2^{3/4}}{\cal F}_2(\tilde{s})\\
	-\frac{\sqrt{\pi}c}{2^{3/4}}{\cal F}_1(\tilde{s})\\
	0
\end{array}
\right)~.
\end{equation}

	\begin{figure}[htb]\centering
{\includegraphics[width=0.362\linewidth]  {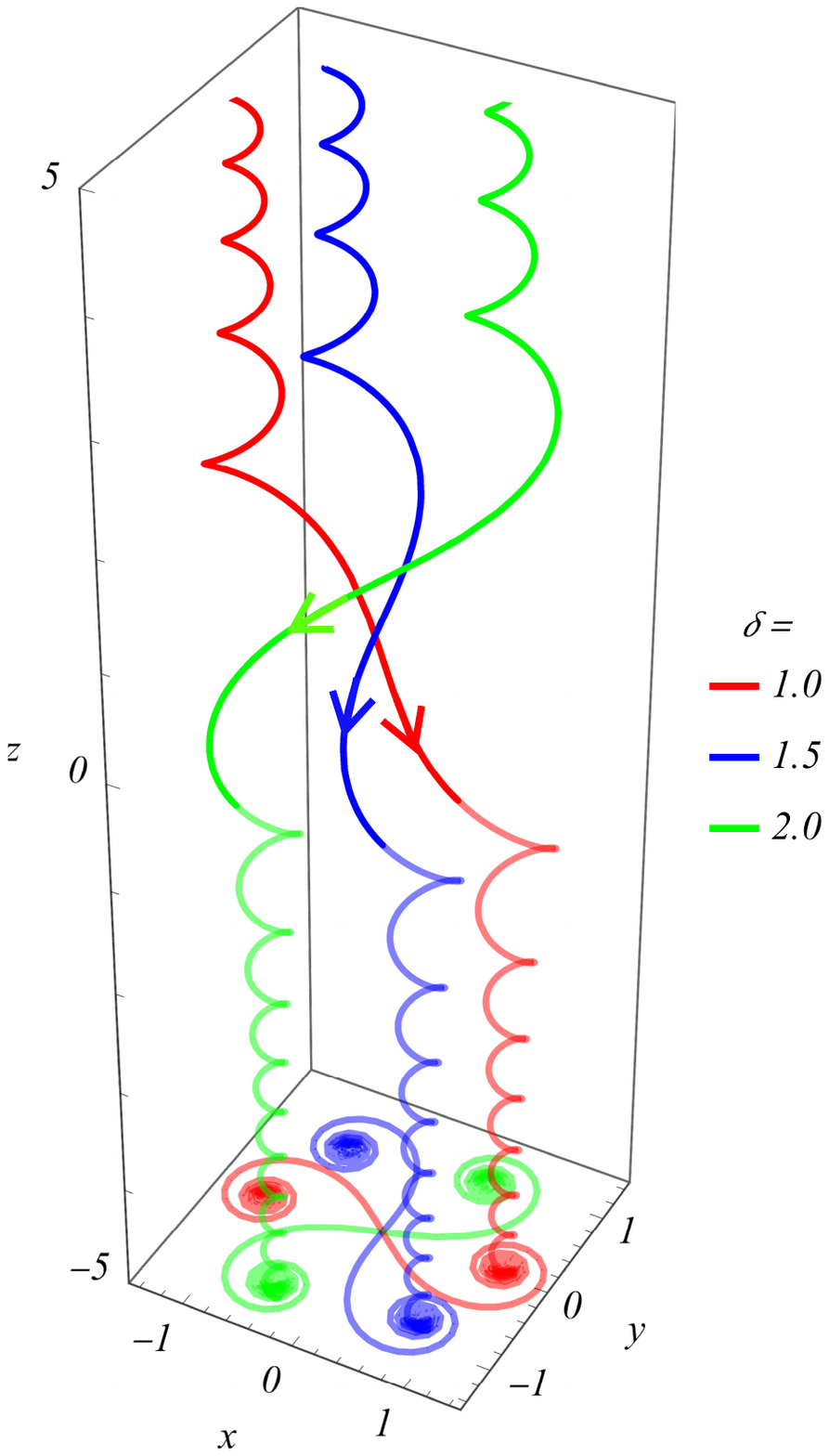}} 
		\caption{$\delta$-shifted clothoid helices $\vec{{\cal C}}_{2,\delta}(\tilde{s})$ for three different values of $\delta$ and $c=1$.}\label{fdeltac2}
	\end{figure}
The coordinates of the $\delta$-foci are $x_{2\delta,F}(\tilde{s}\to\pm\infty)=x_{1\delta,F}(\tilde{s}\to\pm\infty)$ and 
$y_{2\delta,F}(\tilde{s}\to\pm\infty)=-y_{1\delta,F}(\tilde{s}\to\pm\infty)$.                           
The plots of $\vec{{\cal C}}_{2,\delta}(\tilde{s})$ for the same values of the shift parameter as in the first case 
 are displayed in Fig.~\ref{fdeltac2}.

\ms 

The main effect of the shift parameter is a displacement of the inflection point of the helix by an amount $\delta$ in height, which governs the transition region of the helix between the two foci. To illustrate this displacement effect, we exploit the discrete ordered structure arising from the fact that this parameter enters as a phase shift in the Riccati solution.
Therefore, for an ideal shifted clothoid helix, which begins at one focus and terminates at the other, one can obtain a countable infinity of values of $\delta$ from the property of the two foci of being at equal distance with respect to the origin either on the first or the second bisector.
	\begin{equation} \label{focidelta}
\cos\left(\frac{\delta^2}{\sqrt{2}c^2}\right)+\sin\left(\frac{\delta^2}{\sqrt{2}c^2}\right)=\sin\left(\frac{\delta^2}{\sqrt{2}c^2}\right)-
\cos\left(\frac{\delta^2}{\sqrt{2}c^2}\right)
	\end{equation}
yielding
\begin{equation}\label{delta-last}
\delta_n=\pm 2^{1/4}c\sqrt{\frac{(2n+1)\pi}{2}},~~n\in\mathbb{N}~.
\end{equation}

\ms

\begin{figure}[htb]\centering
\subfigure[~ even $n$]{\includegraphics[width=0.49\linewidth]  {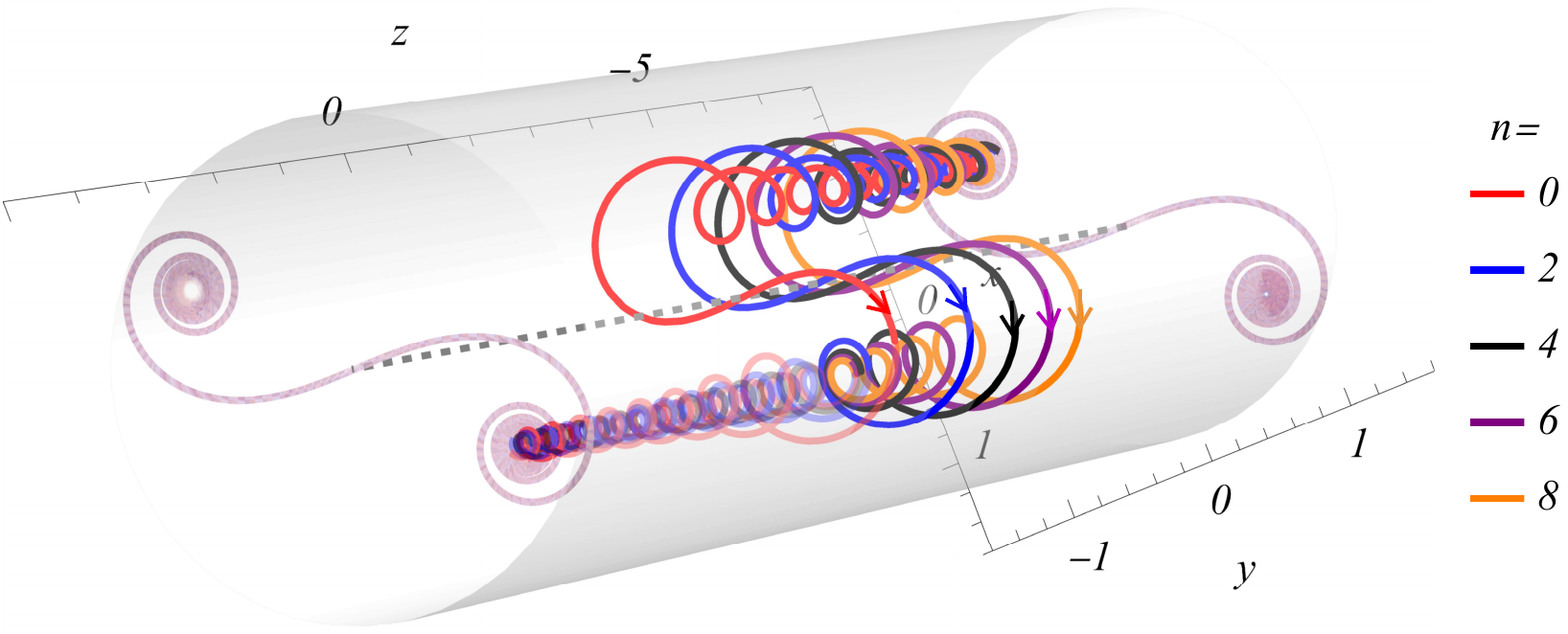}} 
\subfigure[~ odd $\,n$]{\includegraphics[width=0.49\linewidth]  {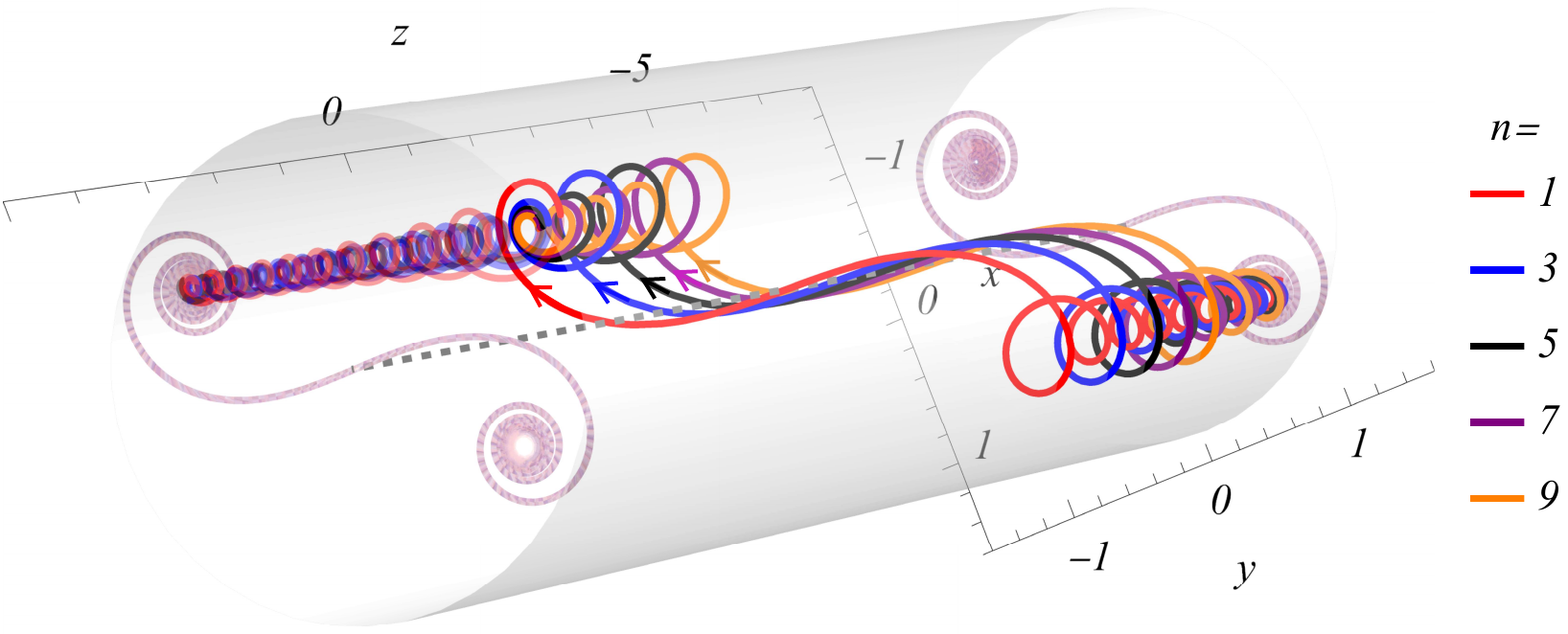}} 
		\caption{$\delta$-clothoid helices $\vec{{\cal C}}_{1,\delta}$ for $k=1, c=1$ and even and odd values of $n$ according to (\ref{delta-last}).}\label{fdeltac3}
	\end{figure}
\begin{figure}[htb]\centering
\subfigure[~ even $n$]{\includegraphics[width=0.49\linewidth]  {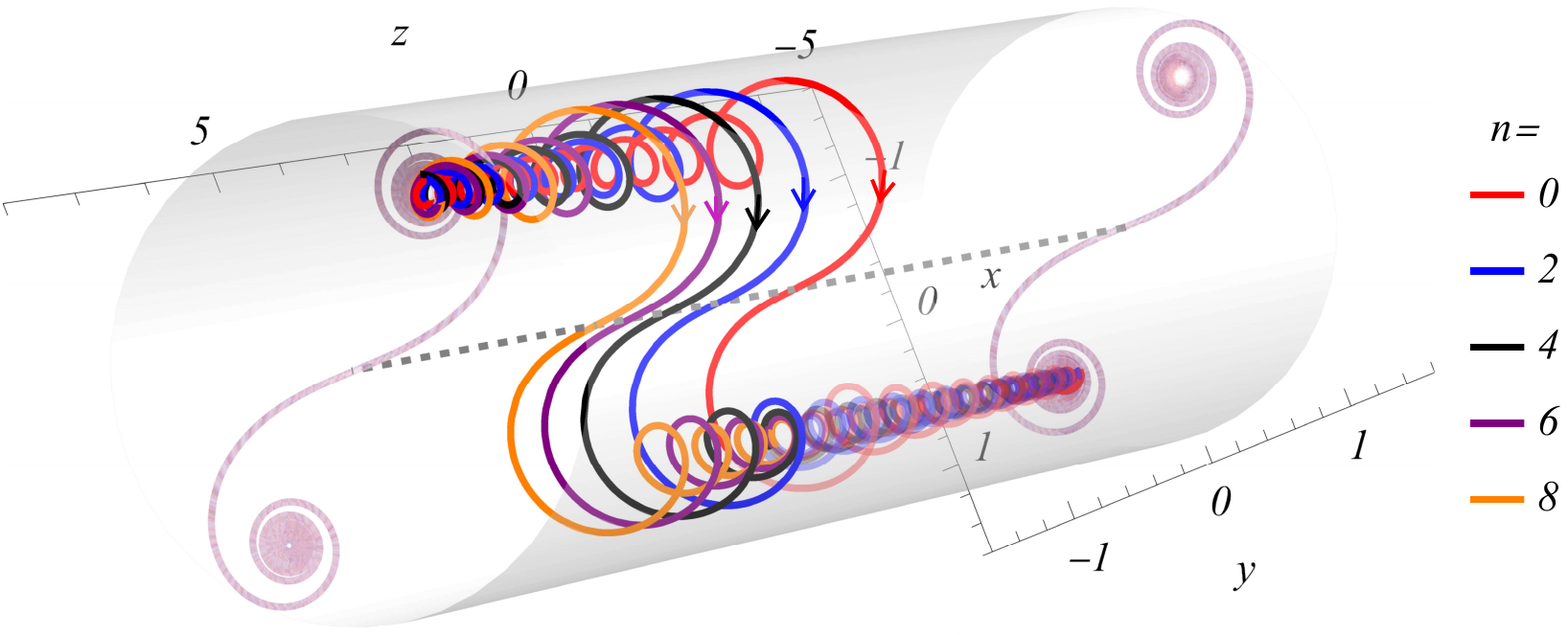}} 
\subfigure[~ odd $n$]{\includegraphics[width=0.49\linewidth]  {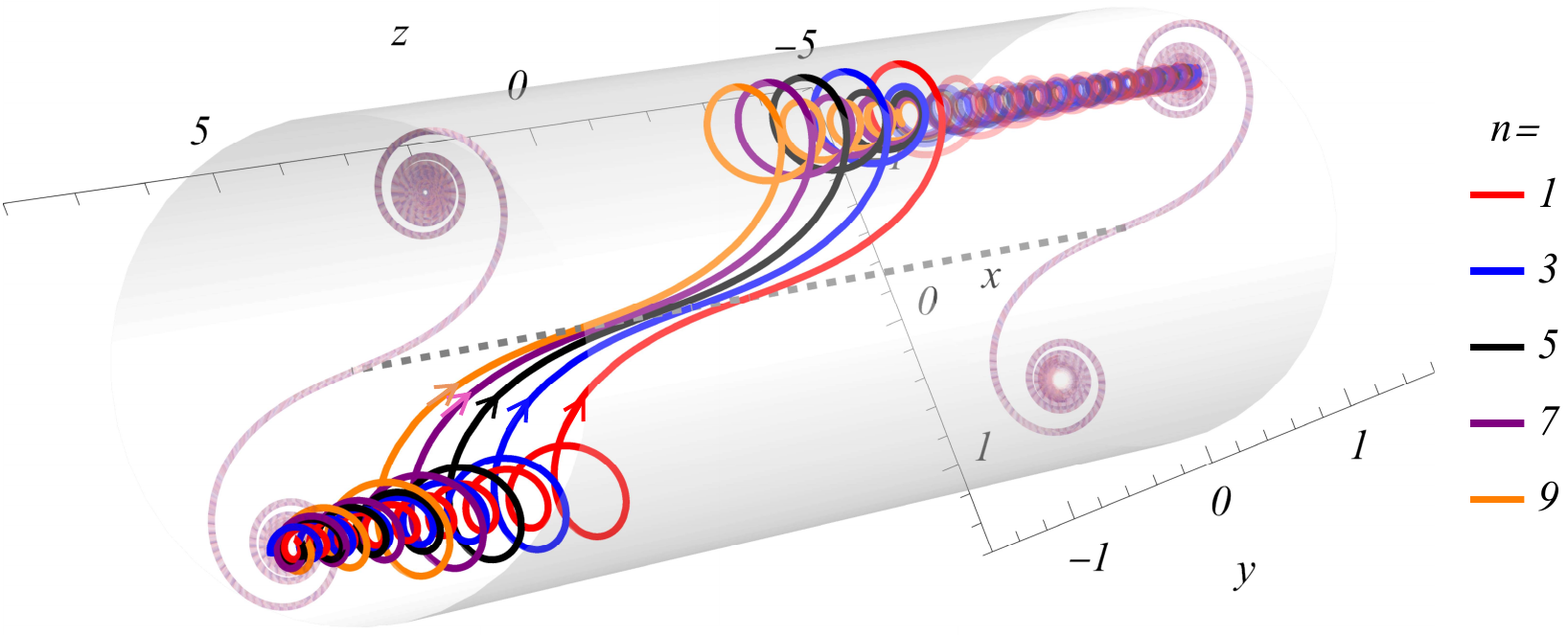}} 
		\caption{Same as in the previous figure for $\delta$-clothoid helices $\vec{{\cal C}}_{2,\delta}$. 
}\label{fdeltac4}
	\end{figure}

Some $\delta$-shifted chlotoid helices of type $\vec{{\cal C}}_{1,\delta}$ and $\vec{{\cal C}}_{2,\delta}$ for different $\delta_n$'s are displayed in Figs.~\ref{fdeltac3} and \ref{fdeltac4}, respectively.   


\newpage 

 \section{Conclusions}

 Clothoid helices, including their phase-shifted counterparts, have been derived via the Lie-Darboux method and some of their geometric features have been analyzed.
 In general, other types of three-dimensional curves can be obtained in this way, but this depends crucially on casting the general solution of the 
Riccati equation in the appropriate rational form.

\ms

The primary applications are expected in optics and acoustics, analogous to, but not limited to the diffraction applications of the corresponding clothoid/Cornu spirals. For example, in photonics, one may envisage structured light beams and pulses with clothoid helical energy density flux 
\cite{Jhajj,QChen} and also of generating three-dimensional optical vortex lattices of this kind behind amplitude transparency masks \cite{Ikon}.  

\ms 


\bigskip

\noindent {\bf Credit author statement}

\medskip

H.C. Rosu: Writing - original draft, Supervision, Formal analysis.

J. de la Cruz: Calculation, Investigation, Formal analysis.

P. Lemus-Basilio: Investigation, Formal analysis.

\bigskip

\noindent {\bf Declaration of Competing Interest}

\medskip

The authors declare that they have no known competing financial interests or personal relationships that could have appeared
to influence the work reported in this paper.

\bigskip

\noindent {\bf Data availability}

\medskip

No data was used for the research described in the paper.

\bigskip
\bigskip

%


\end{document}